\documentstyle[eqsecnum,aps,prb]{revtex}
\begin{document}


\title{Residual resistivity ratio and its relation to the magnetoresistance
behavior in LaMn$_{2}$Ge$_{2}$-derived alloys} 
\author{S. Majumdar, R. Mallik, E.V. Sampathkumaran$^*$ and P.L. Paulose}
\address{Tata Institute of Fundamental Research,  Homi  Bhabha  Road,
Mumbai-400005, INDIA}
\maketitle

\baselineskip 0.6cm

\begin{abstract}
Results of low temperature magnetoresistance ($\Delta\rho/\rho$) and
isothermal magnetization (M) measurements on  polycrystalline $\it
ferromagnetic$ (T$_C$ close to 300 K) natural multilayers,
LaMn$_{2+x}$Ge$_{2-y}$Si$_y$, are reported.  It is found that the samples
with large residual resistivity ratio, $\rho$(300K)/$\rho$(4.2K), exhibit
{\it large positive} magnetoresistance at high magnetic fields.  The
Kohler's rule is not obeyed in these alloys.  In addition, at 4.5 K, there
is a tendency towards linear variation of $\Delta\rho/\rho$ with magnetic
field with increasing $\rho$(300K)/$\rho$(4.2K); however, the field
dependence of $\Delta\rho/\rho$ does not track that of M, thereby
suggesting that the magnetoresistance originates from non-magnetic layers.
It is interesting that these experimental findings on bulk polycrystals
are qualitatively similar to what is seen in artificially grown multilayer
systems recently.
\end{abstract}

\twocolumn
\vskip 2cm
The investigation of magnetic multilayers has been a field of active
interest in view of the giant magnetoresistance behaviour (GMR)
exhibited by these systems.\cite{1} The physics behind electron transport
in these systems is not yet fully understood.\cite{2} Though the magnetic
multilayers typically show \cite{3} large negative magnetoresistance,
$\Delta\rho/\rho$, (defined as [$\rho$(H)-$\rho$(0)]/$\rho$(0) where H is
the applied magnetic field), there have been recent reports of {\it
positive} $\Delta\rho/\rho$ of equally large magnitude in some systems,
e.g. Dy/Sc superlattices\cite{4} with Dy being magnetic, and non magnetic
Cr/Ag/Cr trilayers.\cite{5}  In these systems,  the application of an
external magnetic field modifies the electron scattering/reflection across
the interfaces; hence interfacial contribution to resistivity plays an
important role in determining the $\Delta\rho/\rho$ behavior as evidenced
by higher $\Delta\rho/\rho$ values if the mean free path ($\ell$) is
large.  There is a breakdown of Kohler's rule and there is an inverse
correlation between the residual resistivity ratio (RRR= $\rho$(300
K)/$\rho$(4.2 K)) and large positive magnetoresistance (LPMR) in Ref.  5.
\par
We have earlier reported\cite{6,7} LPMR behavior, qualitatively similar to
the Dy/Sc and Cr/Ag/Cr systems\cite{4,5}, on {\it natural multilayer}
systems, viz., polycrystalline materials with a layered structure.  These
systems are: RMn$_2$Ge$_2$ (R= Sm, La) crystallizing in the
ThCr$_2$Si$_2$-type tetragonal structure\cite{8}, containing atoms in the
layered sequence Th-Si-Cr-Si-Th-Si-Cr-Si-Th along the c-axis. The finding
of note is that the sign of $\Delta\rho/\rho$ is positive at low
temperatures inspite of the fact that the compounds are ferromagnetic with
the Curie temperature around 300 K.  Here, we report the results of our
studies on several annealed and as-cast alloys derived from the compound,
LaMn$_2$Ge$_2$, to look for correlations between RRR and the magnitude as
well as the field dependence of $\Delta\rho/\rho$, seen in those
multilayers.\cite{4,5} We induce a change in RRR by small variations in
Mn-Ge ratio, small chemical substitutions at Mn/Ge sites and in the
sample-preparative conditions.
\par
The polycrystalline samples, LaMn$_{2+x}$Ge$_{2-y}$Si$_y$, mentioned in
table 1 (except the sample 1, which was prepared\cite{7} by arc melting),
were prepared by melting together constituent elements in an induction
furnace. Slight excess of Mn  was added  to compensate for weight loss
while melting. In the present studies, we have adopted an induction
melting method for the preparation of samples since this method is found
to allow better control over Mn loss compared to the arc melting method.
For each sample, a part of the as-cast ingot was homogenised  in  an
evacuated sealed quartz tube at 800 C for 1 week.  X-ray diffraction
patterns (Cu K$_{\alpha }$) confirm single phase nature of the  alloys.
The basal plane parameter is close to 4.2\AA \hskip 0.2cm and the
c-parameter is close to 11\AA \hskip 0.2cm in all the alloys. The
electrical resistivity ($\rho$) measurements were performed by a
conventional four-probe method employing silver paint for making
electrical contacts of the leads with the samples, with the direction of
the current being the same as that of H (longitudinal mode).  The
isothermal magnetization (M) was measured at 4.5 K using a commercial
(Quantum design Inc.) superconducting quantum interference device.
\par
In Table 1, we summarize the values of RRR and $\Delta\rho/\rho$ to
highlight the disorder effect on the $\Delta\rho/\rho$ of the
LaMn$_2$Ge$_2$-derived alloys.  It is interesting to note that the
LaMn$_2$Ge$_2$ sample prepared by induction method (sample 6) is
characterized by a smaller magnetoresistance  compared to the previous
sample prepared by arc melting (sample 1). Thus different preparative
conditions introduce different degree of defects and disorder in the
samples.     It is clear that, as the RRR decreases, the LPMR effect as
seen at 4.5 K is gradually washed out, viz., for RRR $<$ 10, the low
temperature $\Delta\rho/\rho$, though positive, is of small magnitude,
showing that disorder, as inferred from the magnitude of RRR, kills the
LPMR behavior in these systems.  It appears that the ferromagnetism of Mn
layer in LaMn$_2$Ge$_2$ plays no direct role on the magnetoresistance, as
otherwise one would have seen negative magnetoresistance as is usually the
case with polycrystalline ferromagnetic materials;  this is further
corroborated  by a comparison of $\Delta\rho/\rho$ with the isothermal
magnetization behavior at 4.5 K: M saturates for small values of H for all
samples (typical behavior is shown for a few samples in Fig.  2c).  Thus,
as in the case of Dy/Sc multilayer,\cite{4} M and $\Delta\rho/\rho$ do not
track each other.  At this juncture, it may be stated that the electrical
conduction may be favored along the basal plane as indicated by single
crystal investigations\cite{9} on SmMn$_2$Ge$_2$. However, the
non-magnetic layers, but not the Mn layer, are the ones which contribute
to magnetic response from the above arguments.  The fact that we have
observed\cite{10} LPMR even in nonmagnetic alloys belonging to the same
structure if RRR is large, supports the idea that the magnetic Mn layer is
not responsible for the observed behavior. However, {\it the choice of a
ferromagnetic layered compound} enables us to draw {\it definitive}
conclusions on the {\it selective role} of nonmagnetic layers on the
magnetoresistance behavior; though we have seen large values in layered
antiferromagnets in the past,\cite{11} such a conclusion was not possible
from those data, as antiferromagnetism can also give rise to positive
values. We further speculate that the non-magnetic layer acting as an
interface is the one that contains Ge, since low temperature positive
magnetoresistance could be observed even in the ferromagnetic, isomorphous
Sm sample in which Sm layer is ferromagnetic at 4.2K.\cite{6}
\par
In Fig. 1b and 1d we show the temperature dependence of $\Delta\rho/\rho$,
derived from the data shown in Fig 1a and 1c respectively. The values of
$\ell$ have also been evaluated from the $\rho$ data from the knowledge of
the unit-cell density (6.8 gm/cc) and the number of conduction electrons
(estimated to be 19, assuming tri, tetra, and tetra positive La, Mn and
Si/Ge respectively) within the Sommerfeld theory of conduction.\cite{12}
The assumption of theoretical mass density and uncertainties in the
absolute determination of $\rho$, for instance, due to microcracks in the
sample,  and in the number of conduction electrons  introduce errors in
the derived values of $\ell$. Therefore, not much significance may be
attached to these values of $\ell$.  For qualitative arguments,
$\ell$ thus obtained for the two most magnetoresistive samples are shown
in the figures 1b and 1d (continuous lines through the data points).  The
point to be noted is that the $\ell$ closely follows $\Delta\rho/\rho$ as
a function of temperature. It may be remarked that the value of $\ell$ at
which significant positive $\Delta\rho/\rho$ starts appearing is above
3\AA. We derived the same cut-off limit from the low temperature $\rho$
data of all the samples (see table 1) as well. Therefore, though the
absolute values of $\ell$ may not be reliable, on this basis, one can draw
a conclusion that there is a close relationship between the magnitude of
the magnetoresistance and $\ell$.  However,  for the reason mentioned
above, we refrain from drawing more conclusions from the derived $\ell$
values.
\par 
In Fig.  2a and 2b, we present the results of $\Delta\rho/\rho$ as a
function of H at 4.5 K for some of the samples which exhibit significant
magnetoresistance. It is to be remarked that the plot of $\Delta\rho/\rho$
versus H/$\rho$  is not a single line (shown  for one sample in Fig. 3)
and thus the Kohler's rule is found to break down - this is similar to the
observations in Cr/Ag/Cr trilayers{\cite 5}. For the most magnetoresistive
alloy (sample 1), the field dependence is linear in the entire field range
of measurement, instead of a classical quadratic dependence.\cite{13} For
samples 2 and 3, we observe a deviation from this linearity  at low fields
(shown in the inset of Fig. 2a for sample 2) beyond which there is a
linear behaviour.  The low field quadratic dependence could be observable
for samples 3 and 4 with still smaller RRR values. Thus there is a
tendency towards classical quadratic dependence with a decrease of RRR.
(In one of the samples, LaMn$_{1.9}$Ge$_{2.1}$ (I) in Fig. 2b, we see a
convex curvature around a higher field of 20 kOe, the reason for which is
unclear). Thus, the low field linear dependence of magnetoresistance
appears to be a signature of relatively better crystallographic order in
these alloys. In the same way, in the Cr/Ag/Cr multilayers{\cite 5}, an
increase in  RRR brings about a transition from quadratic to linear
behavior in the $\Delta\rho/\rho$ vs H plots, in addition to increasing
the magnitude of observed $\Delta\rho/\rho$. At this juncture, we would
like to recall a recent report\cite{14} on nonmagnetic bulk silver
chalcogenides of the type Ag$_{2+\delta}$(Se,Te): For stoichiometric
($\delta$= 0) samples, there is no significant magnetoresistance; however,
possible {\it introduction of electronic defects} for $\delta$ $>$ 0 gives
rise to LPMR behavior. In these systems also, the magnetoresistance is
found to exhibit an almost linear dependence on the applied field.  Our
samples being magnetic, the following point may have to be considered in
addition to various other explanations which can be offerred to the above
non-magnetic systems; that is, in large RRR alloys, the internal magnetic
fields due to ferromagnetism of Mn is large enough that one is well above
the quadratic region already in zero external field; however, the
deviation from the linearity in smaller RRR samples may possibly be due to
randomness of this local field caused by disorder.\cite{15}
\par
In conclusion, we have brought out the effect of residual resistivity
ratio on the magnitude of the magnetoresistance and its dependence on H at
low fields in LaMn$_2$Ge$_2$-derived alloys, resembling those of some
multilayers.\cite{4,5} The expected negative sign of $\Delta\rho/\rho$ is
not seen in any of these ferromagnetic alloys and this observation,
alongwith the isothermal magnetization behavior which is different from
that of $\Delta\rho/\rho$, is taken as the evidence for selective magnetic
field response from the non-magnetic layers. The present results appear to
render an experimental support to a proposal\cite{9} that these layered
compounds may serve as model systems for artificial multilayers. It is
however not clear whether there are unusual domain wall effects in such
ferromagnetic alloys, an aspect which is yet to be settled even in simple
ferromagnetic systems.\cite{16} In this sense, it is worthwhile to study
the $\Delta\rho/\rho$ behaviour on clean single crystal samples of
RMn$_2$Ge$_2$(Si$_2$); this will also enable to comfortably probe the
magnetoresistance in a geometry with the current perpendicular to the
plane, an important investigation difficult to make on
multilayers.\cite{17} In addition, the studies on artificially grown
layers of the constituent elements in the sequence present in the crystal
structure may be rewarding.   We hope that this work will motivate further
investigations of such natural multilayer systems.

\begin{table}
\caption{
Some data on the transport behavior of alloys derived from LaMn$_2$Ge$_2$.
The samples are arranged in decreasing order of residual resistance ratio
(RRR) = $\rho$(300 K)/$\rho$(4.2 K).  $\Delta\rho/\rho$ is the
magnetoresistance (all positive) evaluated at 4.5 K, 50 kOe field. A and I
stand for 'annealed' and 'as-cast' respectively. All samples were
prepared by induction melting except the sample 1.}

\begin{tabular}{c|lcccc}
Sr. no.& Sample & A/I & RRR &  $\Delta\rho/\rho$ (\%)   \\ \hline
1& LaMn$_2$Ge$_2$  (Arc)         & A &   100        & 70   \\
2& LaMn$_{2.2}$Ge$_{1.8}$        & A &   45         & 23   \\
3& LaMn$_2$Ge$_{1.9}$Si$_{0.1}$  & A &   34         & 9    \\
4& LaMn$_{1.9}$Ge$_{2.1}$        & I &   13         & 8    \\
5& LaMn$_{2.2}$Ge$_{1.8}$        & I &   12         & 2.5  \\
6& LaMn$_2$Ge$_2$                & A &   10         & 1.5  \\
7& LaMn$_2$Ge$_{1.8}$Si$_{0.2}$  & I &   9          & $<$1 \\
8& LaMn$_{1.8}$Ge$_{2.2}$        & A &   8          & $<$1 \\
9& LaMn$_{1.9}$Ge$_{2.1}$        & A &   7          & 2    \\
10& LaMn$_2$Ge$_{1.9}$Si$_{0.1}$ & I &   4          & $<$1 \\ 
11& LaMn$_{2.1}$Ge$_{1.9}$       & A &   4          & $<$1 \\
12& LaMn$_{1.8}$Ge$_{2.2}$       & I &   2          & $<$1 \\
13& LaMn$_{2.1}$Ge$_{1.9}$       & I &   2          & $<$1 \\
14& LaMn$_2$Ge$_{1.8}$Si$_{0.2}$ & A &   2          & $<$1 \\

\end{tabular}
\end{table}

\begin{figure}
\caption{
The electrical resistivity ($\rho$) of samples 1 and 2 (see Table 1) as a
function of temperature (4.2-60 K) in the presence and in the absence of a
magnetic field of 50 kOe is plotted in Fig. (a) and (c) respectively.  The
magnetoresistance $(\Delta \rho /\rho )$ obtained from the data shown in
figure (a) and (c) is plotted in (b) and (d) respectively. 
The mean free path values calculated from the $\rho$ data are shown by
continuous lines in (b) and (d).}
\end{figure}

\begin{figure}
\caption{ 
The magnetoresistance $(\Delta \rho /\rho )$ as a function of magnetic
field at 4.5 K for some of the LaMn$_2$Ge$_2$-derived samples (see Table
1).  The lines through the  data  points  serve  as   guides to the eyes.
The lower part shows typical isothermal magnetization behavior at 4.5 K
for two samples.}
\end{figure}
\begin{figure}
\caption
{Kohler's plot for the most magnetoresistive sample (see Table 1)}
\end{figure}
\end{document}